\documentstyle[12pt]{article}
\begin{document}
\title{On the Connection between Generalized Hypergeometric Functions
and Dilogarithms}
\author{{\bf M.A. Sanchis-Lozano} \\
\\
\it Departamento de F\'{\i}sica Te\'orica \\
\it and \\
\it Instituto de F\'{\i}sica Corpuscular ( IFIC ) \\
\it Centro Mixto Universidad de Valencia-CSIC \\
\\
Dr. Moliner 50, 46100 Burjassot, Valencia (Spain) }
\maketitle
\vspace*{0.1in}
\abstract{Several integrals involving powers and ordinary hypergeometric
 functions are rederived by means of a generalized
hypergeometric function of two variables (Appell's function) recovering
some well-known expressions as particular cases. Simple connections between
 dilogarithms and a kind of Appell's function
are shown. A relationship is generalized to polylogarithms.
\vspace{1.5in}
\par
\hspace{10.5cm}
FTUV : 95/49
\par
\hspace{10.5cm}
IFIC : 95/51}
\newpage
Hypergeometric functions play an important role in mathematical physics
since they are related to a wide class of special functions appearing in a
large
variety of fields. In particular, it is well-known a long time ago that
integrals emerging from loop calculations in Feynman diagrams can be written
in terms of hypergeometric
functions \cite{uno}. More recently, generalized hypergeometric functions
of one or several variables have been used in the evaluation of scalar
one-loop Feynman integrals \cite{un} or multiloop ones \cite{berend}.
\par
In this work we firstly rederive \footnote{An exhaustive set of integrals
involving generalized Gauss functions containing ours as a particular
case can be found in \cite{exton}.}
an integral expression involving two ordinary Gauss' functions yielding
a generalized hypergeometric function of two variables (Appell's function).
Several formulae appearing in standard tables ({\em e.g.} Gradshteyn and
Ryzhik \cite{dos}) of utility for the evaluation of Feynman loop integrals
are obtained as particular cases. Moreover, we have shown a simple relationship
between a kind of Appell's function and dilogarithms \cite{lewin},
contributing to enlarge the knowledge on the connection between them.
In the appendices at the end
of the paper we present a brief survey on the generalized Gauss' functions
establishing the notation employed and revising some of their
properties needed in this \vspace{0.3in} work.\newline
{\bf expression 1}
\vspace{0.1in}
\[ \int_{0}^{1}du\ u^{\gamma-1}{(1-u)}^{\rho-1}\
_2F_1[\sigma,\eta;\gamma;zu]
\ _{2}F_{1}[\alpha,\beta;\rho;k(1-u)]= \]
\begin{equation}
=\frac{\Gamma(\gamma)\Gamma(\rho)}{\Gamma(\gamma+\rho)}
\ F_3[\alpha,\sigma,\beta,\eta;\gamma+\rho;k,z]
\end{equation}
\vspace{0.05in}
provided that $Re(\gamma)>0$, $Re(\rho)>0$, ${\mid}arg(1-k){\mid}<\pi$
,\vspace{0.1in} ${\mid}arg(1-z){\mid}<\pi$.\newline
{\em Proof.} We will first show that Eq. (1) holds in the domain of
convergence of the series. Expanding one of the two $_{2}F_{1}$ functions
as a power series leads to:
\[ \sum_{n=0}^{\infty}\frac{(\alpha)_{n}(\beta)_{n}}{(\rho)_{n}}\frac{k^n}{n!}
\int_{0}^{1}du\ u^{\gamma-1}(1-u)^{\rho+n-1}\
_2F_1[\sigma,\eta;\gamma;zu]\ \ \ ;\
{\mid}k{\mid}<1,\ {\mid}z{\mid}<1 \]
where we have interchanged the order of summation and integration
on account of the dominated convergence theorem of Lebesgue, provided that
$Re(\gamma)>0$, $Re(\rho)>0$. Now, performing the integration over $u$ one
gets from (A.3):
\[ \Gamma(\gamma)\sum_{n=0}^{\infty}\frac{(\alpha)_{n}(\beta)_{n}}{(\rho)_{n}}
\frac{k^n}{n!}\frac{\Gamma(\rho+n)}{\Gamma(\gamma+\rho+n)}
\ _{3}F_{2}[\gamma,\sigma,\eta;\gamma+\rho+n,\gamma;z]= \]
\[ =\Gamma(\gamma)\sum_{n=0}^{\infty}\frac{(\alpha)_n(\beta)_n}{(\rho)_n}
\frac{k^n}{n!}\frac{\Gamma(\rho+n)}{\Gamma(\gamma+\rho+n)}
\ _{2}F_{1}[\sigma,\eta;\gamma+\rho+n;z] \]
where a cancellation between two parameters in the $_3F_2$ function
occurred.
\par
Finally, using that: $\Gamma(\rho+n)=\Gamma(\rho)(\rho)_n$ ,
$\Gamma(\gamma+\rho+n)= \Gamma(\gamma+\rho)(\gamma+\rho)_n$, one arrives at
\[ \frac{\Gamma(\gamma)\Gamma(\rho)}{\Gamma(\gamma+\rho)}
\sum_{n=0}^{\infty}\frac{(\alpha)_n(\beta)_n}
{(\gamma+\rho)_{n}}\frac{k^n}{n!}\
_{2}F_{1}[\sigma,\eta;\gamma+\rho+n;z] \]
which leads to Eq. (1) at once in virtue of (B.2). Moreover, the integral (1)
furnishes a single-valued function of two variables beyond the domain
of convergence of the series
by imposing the cuts: ${\mid}arg(1-k){\mid}<\pi$,
${\mid}arg(1-z){\mid}<\pi$ \vspace{0.2in}.
\newline
{\bf expression 1.1}
\par
\vspace{0.1in}
Setting $\rho=\delta-\gamma$, $\alpha=\delta-\sigma$ and $\beta=\delta-\eta$
in Eq. (1) the formula $7.512.7$ of reference \cite{dos}
is recovered: \footnote{Except the exponent of
$(1-k)$ which in our notation would read: $2\sigma-\delta$. Clearly this is
an error since the result should be invariant under the
interchange of $\sigma$ and $\eta$, as the l.h.s. certainly is. The
original source \cite{dosp} is equally wrong.}
\[ \int_{0}^{1}du\ u^{\gamma-1}{(1-u)}^{\delta-\gamma-1}\ _2F_1[\sigma,\eta;
\gamma;zu]\
_{2}F_{1}[\delta-\sigma,\delta-\eta;\delta-\gamma;k(1-u)]= \]
\begin{equation}
=\frac{\Gamma(\gamma)\Gamma(\delta-\gamma)}{\Gamma(\delta)}\
(1-k)^{\sigma+\eta-\delta}\ _2F_1[\sigma,\eta;\delta;k+z-kz]
\end{equation}
provided that $Re(\delta)>Re(\gamma)>0$, ${\mid}arg(1-k){\mid}<\pi$
, ${\mid}arg(1-z){\mid}<\pi$.\par
\vspace{0.1in}
This can be directly obtained from Eq. (1) taking further into account the
property (B.6) which here implies:
\begin{center}
$F_3[\delta-\sigma,\sigma,\delta-\eta,\eta;\delta;k,z]=
(1-k)^{\sigma+\eta-\delta}\ _2F_1[\sigma,\eta;\delta;k+z-kz]$
\end{center}
\vspace{0.2in}
{\bf expression 1.2}
\par
\vspace{0.1in}
With the aid of (A.4) the left hand side of Eq. (1) can be written as:
\vspace{0.1in}
\[ \int_{0}^{1}du\ u^{\gamma-1}{(1-u)}^{\rho-1}{(1-k(1-u))}^{-\alpha}\
_2F_1[\sigma,\eta;\gamma;zu]
\ _{2}F_{1}\biggl[\alpha,\rho-\beta;\rho;\frac{k(1-u)}{(k-1-ku)}\biggr] \]
Now, let us assume that $z$ and $k$ are related through $k=z/(z-1)$. Then
\[ (1-z)^{\alpha}\int_{0}^{1}du\ u^{\gamma-1}{(1-u)}^{\rho-1}
{(1-zu)}^{-\alpha}\ _2F_1[\sigma,\eta;\gamma;zu]
\ _{2}F_{1}[\alpha,\rho-\beta;\rho;\frac{z(1-u)}{1-zu}]= \]
\[ =\frac{\Gamma(\gamma)\Gamma(\rho)}{\Gamma(\gamma+\rho)}\
F_3[\alpha,\sigma,\beta,\eta;\gamma+\rho;z/(z-1),z] \]
Next, let us suppose further that $\beta=\gamma+\rho-\eta$. Then taking
into account consecutively the properties (B.5) and (B.4) the
right hand side of the last expression becomes:
\[ \frac{\Gamma(\gamma)\Gamma(\rho)}{\Gamma(\gamma+\rho)}\ (1-z)^{\alpha}
F_1[\eta;\sigma;\alpha;\gamma+\rho;z,z]
=\frac{\Gamma(\gamma)\Gamma(\rho)}{\Gamma(\gamma+\rho)}\ (1-z)^{\alpha}\
_2F_{1}[\sigma+\alpha,\eta;\gamma+\rho;z] \]
Hence one recovers the formula $7.512.8$ of ref. \cite{dos}:
\[ \int_{0}^{1}du\ u^{\gamma-1}(1-u)^{\rho-1}(1-zu)^{-\alpha}\
_2F_1[\sigma,\eta;\gamma;zu]\
_2F_1\biggl[\alpha,\eta-\gamma;\rho;\frac{z(1-u)}{(1-zu)}\biggr]= \]
\begin{equation}
=\frac{\Gamma(\gamma)\Gamma(\rho)}{\Gamma(\gamma+\rho)}\
_2F_1[\sigma+\alpha,\eta;\gamma+\rho;z]
\end{equation}
provided that $Re(\gamma)>0$, $Re(\rho)>0$, ${\mid}arg(1-z){\mid}<\pi$.
\par
\vspace{0.2in}
Let us now go back again to Eq. (1) and consider $\eta=\gamma$ as a new
special case. Then two parameters of a hypergeometric function
in the integrand cancel, i.e. $_2F_1[\sigma,\gamma;\gamma;zu]=\
_1F_0[\sigma;zu]=(1-zu)^{-\sigma}$, \vspace{0.2in} yielding:
\newline
{\bf expression 2}
\par
\vspace{0.1in}
\[ \int_{0}^{1}du\ u^{\gamma-1}{(1-u)}^{\rho-1}(1-zu)^{-\sigma}
\ _{2}F_{1}[\alpha,\beta;\gamma;ku]= \]
\begin{equation}
=\frac{\Gamma(\gamma)\Gamma(\rho)}{\Gamma(\gamma+\rho)}\ (1-z)^{-\sigma}
F_3[\alpha,\sigma,\beta,\rho;\gamma+\rho;k,z/(z-1)]
\end{equation}
\vspace{0.05in}
provided that $Re(\gamma)>0$, $Re(\rho)>0$,
${\mid}arg(1-k){\mid}<\pi$, \vspace{0.1in} ${\mid}arg(1-z){\mid}<\pi$.\newline
{\em Proof.} It follows inmediately as a particular case of the expression
{\bf 1} by means of the change of the integration variable: $u{\rightarrow}1-u$
and interchanging the $\gamma$ and $\rho$ parameters.\par
An alternative (direct) proof is achieved with the aid of the
integral representation of the $F_3$ Appell's function. Starting from (B.3) and
making the consecutive changes of
the integration variables: $v{\rightarrow}1-v$ and $u{\rightarrow}uv$
it follows that
\[ \frac{{\Gamma}(\beta){\Gamma}(\rho){\Gamma}({\gamma}-\beta)}
{{\Gamma}({\gamma}+\rho)}\
F_3[\alpha,\sigma,\beta,\rho;\gamma+\rho;k,z/(z-1)]
\ = \]
\[ (1-z)^{\sigma}
\int_{0}^{1}\int_{0}^{1}dv\ du\ v^{\gamma-1}u^{\beta-1}(1-v)^{\rho-1}
(1-u)^{\gamma-\beta-1}(1-zv)^{-\sigma}(1-kvu)^{-\alpha} \]
\par
Hence the expression {\bf 2} is immediately obtained by expressing
$_2F_1[\alpha,\beta;\gamma;kv]$ in its Euler's integral representation
\vspace{0.2in} (A.2).
\newline
{\bf expression 2.1}
\par
\vspace{0.1in}
Setting $k=1$ in expression {\bf 2} reproduces the result $7.512.9$
of ref. \cite{dos}:
\[ \int_{0}^{1}du\ u^{\gamma-1}{(1-u)}^{\rho-1}{(1-zu)}^{-\sigma}
\ _{2}F_{1}[\alpha,\beta;\gamma;u]= \]
\[ =\frac{\Gamma(\gamma)\Gamma(\rho)\Gamma(\gamma+\rho-\alpha-\beta)}
{\Gamma(\gamma+\rho-\alpha)\Gamma(\gamma+\rho-\beta)}(1-z)^{-\sigma}
\ _3F_2[\rho,\sigma,\gamma+\rho-\alpha-\beta;\gamma+\rho-\alpha,
\gamma+\rho-\beta;z/(z-1)] \]
provided additionally that: $Re(\gamma+\rho-\alpha-\beta)>0$.\par
\vspace{0.1in}
This can be easily shown by rewriting the power
expansion of $F_3$ following (B.2) in terms of
$_2F_1[\alpha,\beta;\gamma+\rho+n;1]$ supposed the convergence of the
series, and using the Gauss' summation relation:
\[ \ _2F_1[a,b;c;1]=\frac{\Gamma(c)\Gamma(c-a-b)}
{\Gamma(c-a)\Gamma(c-b)}\ \ \ \ \ ;\ Re(c-a-b)>0 \]
\par
\vspace{0.3in}
\subsubsection*{Particular values of the parameters of $F_3$}
Let us now take the particular values of the parameters:
$\alpha=\beta=\sigma=\eta=\gamma=1$ and $\rho=2$ in the
$F_3[\alpha,\sigma,\beta,\eta;\gamma+\rho;x,y]$ Appell's function. From
our expression {\bf 1}, it is easy to see that making $y=1$ ones gets
\begin{equation}
x\ F_3[1,1,1,1;3;x,y=1]\ =\ 2x\ _3F_2[1,1,1;2,2;x]\ =\ 2\ Li_2(x)
\end{equation}
where Eq. (A.8) has been taken into account. In fact, one can get the same
result by expanding $F_3[1,1,1,1;3;x,1]$ in terms of $_2F_1[1,1;3+m;1]$ and
using the Gauss' summation relation. (See also appendix B for a generalization
to polylogarithms.) The restriction ${\mid}x{\mid}<1$ can be dropped on
account of analytic continuation, extending the domain of analyticity over the
complex $x$-plane cut from $1$ to $\infty$ along the real axis. It is obvious
from symmetry, that an equivalent expression for $y$ must be satisfied. In
fact, Eq. (5) can be viewed as a particular case of a more
general relationship between this Appell's series and \vspace{0.2in}
dilogarithms:
\newline
{\bf expression 3}
\par
\vspace{0.1in}
\begin{equation}
\frac{1}{2}xy\ F_{3}[1,1,1,1;3;x,y]=Li_{2}(x)+Li_{2}(y)-Li_{2}(x+y-xy)
\end{equation}
\vspace{0.1in}
${\mid}arg(1-x){\mid}<\pi$, \vspace{0.1in} ${\mid}arg(1-y){\mid}<\pi$.\newline
{\em Proof.} This formula can be again derived from expression
{\bf 1} by calculating directly the integral in terms of dilogarithms.
Instead, we will prove it by
differentiating both sides with respect
to $x$ and $y$ consecutively. Expanding the Appell's function as a double
series, the result of differentiating the l.h.s. reads:
\begin{equation}
\frac{1}{2}\ F_3[1,2,2,1;3;x,y]
\end{equation}
Now, invoking the property (B.6):
\begin{center}
$F_3[\alpha,\gamma-\alpha,\beta,\gamma-\beta;\gamma;x,y]=
(1-y)^{\alpha+\beta-\gamma}\ _2F_1[\alpha,\beta;\gamma;x+y-xy]$
\end{center}
which is valid in a suitable small open polydisc centered at the origin, we
conclude that (7) can be rewritten as
\[ \frac{1}{2}\ _2F_1[1,2;3;x+y-xy] \]
\par
Next, differentiating twice the r.h.s of Eq. (6) one gets:
\[ -\frac{1}{x+y-xy}[1+\frac{1}{x+y-xy}\ln{(1-(x+y-xy))}]\ =\
\ \frac{1}{2}\ _2F_1[1,2;3;x+y-xy] \]
the last step coming from (A.5). Then both sides in Eq. (6) would differ
in $f(x)+g(y)$:
\[ \frac{1}{2}xy\ F_{3}[1,1,1,1;3;x,y]=Li_2(x)+Li_2(y)-Li_{2}(x+y-xy)+
f(x)+g(y) \]
where $f(x)$ and $g(y)$ are functions to be determined by taking particular
values of the variables. Setting $x=0$ and $y=0$
it is easy to see that $f(x)=g(y){\equiv}0$.\par
Now, by analytic continuation we dispense with the restriction on the small
polydisc, extending its validity to a suitable domain of $C^2$:
in order to get a single-valued function, with a well-defined branch
for each dilogarithm in Eq. (6), we assume further that
${\mid}arg(1-x){\mid}<\pi$, \vspace{0.2in}
${\mid}arg(1-y){\mid}<\pi$. \footnote{Observe that
then each function of one complex variable obtained from (6) by fixing the
other
variable is analytic in the corresponding subset of $C^2$. Then the
function of two variables $\frac{1}{2}xyF_3$ is analytic according to the
theorem of Hartogs-Osgood}\newline
{\bf expression 3.1}
\par
\vspace{0.1in}
\begin{equation}
x^2\ F_{3}[1,1,1,1;3;x,-x]\ =\ Li_2(x^2)
\end{equation}
for $x$ \vspace{0.1in} real.
\newline
{\em Proof.} It follows directly from Eq. (6) using the
relation $Li_2(x)+Li_2(-x)=\frac{1}{2}Li_2(x^2)$ \vspace{0.2in}.\newline
{\bf expression 3.2}
\par
\vspace{0.1in}
\begin{equation}
x^2\ F_{3}[1,1,1,1;3;x,x]\ =\ 4\ Li_2\biggl(\frac{1}{2-x}\biggr)\ +\
2\ln{}^2(2-x)\ -\ \frac{{\pi}^2}{3}
\end{equation}
for $x$ real and less than \vspace{0.1in} unity.
\newline
{\em Proof.} It follows directly from Eq. (6) using the relation:
$Li_2(2x-x^2)=2Li_2(x)-2Li_2(1/(2-x))+{\pi}^2/6-\ln{}^2(2-x)$
\vspace{0.2in} \cite{lewin}.
\newline
{\bf expression 3.3}
\par
\vspace{0.1in}
\begin{equation}
lim_{y{\rightarrow}0}\ \frac{xy}{2}\ F_{3}[1,1,1,1;3;x,y]\ =\
y\ \biggl[\ 1\ +\ \frac{1-x}{x}\ln{(1-x)}\ \biggr]
\end{equation}
{\em Proof.} It follows directly from Eq. (6) using the relation:
$_2F_1[1,1;3;x]=(1-x)^{-1}\ _2F_1[1,2;3;x/(x-1)]$ and (A.5). If besides
$x{\rightarrow}0$, the limit $xy/2$ is quickly
\vspace{0.2in} recovered.
\par
The set of expressions {\bf 3} provide new connections (not shown in
literature to our knowledge) between
dilogarithms and a certain $F_3$ Appell's function.
\newpage
\appendix
\renewcommand{\theequation}{\thesection.\arabic{equation}}
\section*{Appendices}
\section{}
{\em Generalized Gauss' Functions of one variable}
\setcounter{equation}{0}
\par
\vspace{0.05in}
Hypergeometric functions can be introduced
at first as series within a certain domain of convergence \cite{cinc}
\cite{six} \cite{seis}. We write, using the abbreviate notation:
\begin{equation}
_{p}F_{q}[\{a\}_p;\{b\}_q;z]\ =\
\sum_{n=0}^{\infty}\frac{(a_1)_n...(a_p)_n}{(b_1)_n...(b_q)_n}\frac{z^{n}}{n!}
\end{equation}
where $(a)_n=\Gamma(a+n)/\Gamma(a)$ stands for the Pochhammer symbol. We
suppose that none of the denominator parameters is a negative integer
or zero. This series
converges for all values of $z$, real or complex, when $p{\leq}q$, and for
${\mid}z{\mid}<1$ when $p=q+1$. In the latter case, it also converges
(absolutely) on the circle ${\mid}z{\mid}=1$ if
$Re\ (\sum_{i=1}^{q}b_{i}-\sum_{i=1}^{p}a_{i})>0$. If $p>q+1$, the series
never converges, except either when $z=0$ or when
the series terminates, that is when one at least of
the $a$ parameters is zero or a negative integer.\par
Hypergeometric series admit in general an integral representation of the
Euler's type \cite{seis} \cite{siet} \cite{nuev}
, which permits the corresponding analytic continuation in the
complex $z$-plane beyond the unit disc.\par
For the ordinary hypergeometric series, we have:
\begin{equation}
_{2}F_{1}[a,b;c;z]\ =\ \frac{\Gamma(c)}{\Gamma(b)\Gamma(c-b)}
\int_{0}^{1}du\  u^{b-1}(1-u)^{c-b-1}(1-zu)^{-a}
\end{equation}
with $Re(c)> Re(b)>0$. In order to get a single-valued analytic function
in the whole complex $z$-plane we will follow the customary convention of
assuming a cut along the real axis from $1$ to $\infty$.\par
For the generalized hypergeometric function of one variable the integral
representation of the Euler's type is:
\begin{equation}
_{p}F_{q}[a_{1},\{a\}_{p-1};b_1,\{b\}_{q-1};z]\ =\
\frac{\Gamma(b_{1})}{\Gamma(a_1)
\Gamma(b_{1}-a_{1})}\int_{0}^{1}du\ u^{a_{1}-1}(1-u)^{b_{1}-a_{1}-1}\
_{p-1}F_{q-1}[\{a\}_{p-1};\{b\}_{q-1};zu]
\end{equation}
under the constraints: $p{\leq}q+1$, $Re(b_1)>Re(a_1)>0$ and none of
$b_i,\ i=1...q$, is zero or a negative integer,
giving the analytic continuation in the whole complex $z$-plane, cut along the
positive axis from $1$ to $\infty$ again.\par
A well-known transformation between
Gauss' hypergeometric functions of one variable, needed in the main text
is: \cite{nuev}
\vspace{0.1in}
\[ \ _2F_1[a,b;c;z]=(1-z)^{c-b-a}\ _2F_1[c-a,c-b;c;z] \]
\begin{equation}
=\ (1-z)^{-a}\ _2F_1[a,c-b;c;z/(z-1)]\ =
\ (1-z)^{-b}\ _2F_1[c-a,b;c;z/(z-1)]
\end{equation}
\par
\vspace{0.05in}
An interesting relation between an ordinary Gauss' function and an
elementary function not usually shown in specialized tables is:
\begin{equation}
z\ _2F_1[1,2;3;z]=-2\ [\ 1+\frac{1}{z}\ln{(1-z)}\ ]
\end{equation}
which can be proved by expanding both sides as power series.
\vspace{0.05in}
\subsubsection*{The dilogarithm and its relation to
the $_3F_2$ function}
\vspace{0.05in}
The dilogarithmic function is defined as: \cite{nue2} \cite{lewin}
\begin{equation}
Li_2(z)=-\int_{0}^{1}\ du\ \frac{\ln{(1-zu)}}{u}
\end{equation}
for values of $z$ real or complex. If ${\mid}z{\mid}<1$, the dilogarithm
may be expanded as the power series:
\begin{equation}
Li_2(z)\ =\ \sum_{n=0}^{\infty}\frac{z^{n+1}}{(n+1)^2}
\end{equation}
corresponding to the principal value. We can also write:
\begin{equation}
 Li_2(z)=z\int_{0}^{1}\ du\ _2F_1[1,1;2;zu]\ =\ z\
_3F_2[1,1,1;2,2;z]
\end{equation}
The derivative of the dilogarithm is
\begin{equation}
\frac{d}{dz}Li_2(z)=-\frac{\ln{(1-z)}}{z}
\end{equation}
\newpage
\section{}
{\em Generalized Gauss' Functions of two variables
: Appell's functions}
\setcounter{equation}{0}
\par
\vspace{0.05in}
In this paper we are involved in particular with the $F_3$ Appell's function
 \cite{dos} \cite{six} \cite{seis} \cite{siet}, so we write its series
expansion:
\begin{equation}
F_3[a,a',b,b';c;x,y]\ =\ \sum_{m=0}^{\infty}\sum_{n=0}^{\infty}
\frac{(a)_{m}(a')_{n}(b)_{m}(b')_{n}}{(c)_{m+n}}\frac{x^m}{m!}\frac{y^n}{n!}
\end{equation}
which exists for all real or complex values of $a, a', b, b',$ and
$c$ except $c$ a negative integer. With
regard to its convergence, the $F_3$ series is absolutely convergent
when both ${\mid}x{\mid}<1$ and ${\mid}y{\mid}<1$. Then there is no
problem with internal rearrangements of the series.\par
The $F_3$ function can be rewritten in terms of ordinary Gauss' functions:
\begin{equation}
F_3[a,a',b,b';c;x,y]\ =\ \sum_{m=0}^{\infty}\frac{(a)_m(b)_m}{(c)_m}
\frac{x^m}{m!}\ _2F_1[a',b';c+m;y]
\end{equation}
where we have made use of the relation: $(c)_{m+n}=(c)_m(c+m)_n$.\par
Moreover, the $F_3$ function admits the following integral representation:
\cite{dos} \cite{seis}
\begin{equation}
F_3[a,a',b,b';c;x,y]=\ \frac{{\Gamma}(c)}{{\Gamma}(b){\Gamma}(b')
{\Gamma}(c-b-b')}\ {\times}
\end{equation}
\[ \int\int\ du\ dv\ u^{b-1}v^{b'-1}(1-u-v)^{c-b-b'-1}(1-xu)^{-a}
(1-yv)^{-a'} \]
where the integral is taken over the triangular region $0{\leq}u$
, $0{\leq}v$, $u+v{\leq}1$, under the conditions: $Re(b)>0$, $Re(b')>0$
, $Re(c-b-b')>0$. This expression furnishes a single-valued analytic function
in the domain defined by the Cartesian product of the complex planes
of $x$ and $y$ with the restrictions
${\mid}arg(1-x){\mid}<\pi$, ${\mid}arg(1-y){\mid}<\pi$. Hence, the
order of integration may be reversed according to Fubini's theorem.\par
Some properties and relations between Appell's functions needed in this paper
are given below: \cite{dos} \cite{seis}
\begin{equation}
F_1[a;b,b';c;x,x]\ =\ _2F_1[a,b+b';c;x]\ =\
_2F_1[b+b',a;c;x]
\end{equation}
\vspace{0.05in}
\begin{equation}
F_3[a,c-a,b,b';c;x,y/(y-1)]=F_3[b',b,c-a,a;c;y/(y-1),x]=
(1-y)^{b'}F_1[a;b,b';c;x,y]
\end{equation}
\vspace{0.05in}
\begin{equation}
F_3[a,c-a,b,c-b;c;x,y]=F_3[c-a,a,c-b,b;c;y,x]=
(1-y)^{a+b-c}\ _2F_1[a,b;c;x+y-xy]
\end{equation}
\newpage
\subsubsection*{The polylogarithm and its relation to the generalized
Camp\'e de F\'eriet function $F_B^{(2)}$}
The polylogarithm $Li_q(z)$ is defined as a series as
\begin{equation}
Li_q(z)\ =\ \sum_{n=0}^{\infty}\frac{z^{n+1}}{(n+1)^q}\ \ \ \ (q>1)
\end{equation}
which can be expressed according to
\[ Li_q(z)=z\int_{0}^{1}\ du\ _qF_{q-1}[\{1\}_q;\{2\}_{q-1};zu]\
=\ z\ _{q+1}F_{q}[\{1\}_{q+1};\{2\}_q;z]  \]
A generalized Camp\'e de F\'eriet function \footnote{Camp\'e de F\'eriet
functions are special cases of generalized Lauricella functions
of two variables \cite{un}}, of particular interest for us, is defined as
\begin{equation}
F_B^{(2)}[\{b\}_r,\{b'\}_s;\{d\}_t;c;x,y]\ =\
\sum_{m=0}^{\infty}\sum_{n=0}^{\infty}
\frac{(b_1)_{m}...(b_r)_{m}(b'_1)_{n}...(b'_s)_{n}}
{(d_1)_m...(d_t)_m\ (c)_{m+n}}\ \frac{x^m}{m!}\frac{y^n}{n!}
\end{equation}
Thus the following equality is satisfied
\begin{equation}
x\ F_B^{(2)}[\{1\}_q,\{1\}_2;\{2\}_{q-2};3;x,y=1]\ =\ 2\ Li_q(x)
\end{equation}
which is the generalization of Eq. (5).
\end{document}